\documentclass[pdflatex,iicol,sn-aps]{sn-jnl}
\usepackage{graphicx}%
\usepackage{physics}
\usepackage{multirow}%
\usepackage{amsmath,amssymb,amsfonts}%
\usepackage{amsthm}%
\usepackage{mathrsfs}%
\usepackage[title]{appendix}%
\usepackage{xcolor}%
\usepackage{textcomp}%
\usepackage{manyfoot}%
\usepackage{booktabs}%
\usepackage{algorithm}%
\usepackage{algorithmicx}%
\usepackage{algpseudocode}%
\usepackage{listings}%
\usepackage{titlesec}

\usepackage[english]{babel}
\usepackage{physics}
\usepackage{graphicx}
\usepackage{bm}
\usepackage{pgfplots}
\usepackage{tikz}
\usepackage{amsmath}
\usepackage{amssymb}
\usepackage{comment}
\usepackage{siunitx}
\usepackage{physics}
\usepackage{xfrac}
\usepackage{relsize}
\usepackage{mleftright}
\newcommand{\ad}[1]{a_{#1}^\dagger}
\newcommand{\an}[1]{a_{#1}\z}

\newcommand{\bn}[1]{b_{#1}^{\phantom\dagger}}
\newcommand{\bd}[1]{b_{#1}^{\dagger}}
\newcommand{\cn}[1]{c_{#1}^{\phantom\dagger}}
\newcommand{\cd}[1]{c_{#1}^{\dagger}}
\newcommand{\ncr}{\nonumber\\}

\newcommand{\coss}[1]{\cos^2\mleft(#1\mright)}

\newcommand{\dm}[1]{\ket{#1}\bra{#1}}

\newcommand{\ed}{\epsilon_d\z}
\newcommand{\eds}{\epsilon_d^2}

\newcommand{\fn}[1]{f_{#1}\z}
\newcommand{\fd}[1]{f_{#1}^\dagger}
\newcommand{\ff}{\mathcal{F}}

\newcommand{\gon}{\mv_{\mbox{on}}\z}
\newcommand{\goff}{\mv_{\mbox{off}\z}}
\newcommand{\hc}{\mathrm{H.~c.}}
\newcommand{\hn}[1]{\mathcal{H}_{#1}\z}
\newcommand{\hnrg}{\mathcal{H}_{\lambda,\zeta}\z}

\newcommand{\mc}[1]{\mathcal{C}_{#1}\z}
\newcommand{\mn}{\mathcal{N}}
\newcommand{\mv}{\mathcal{V}}
\newcommand{\nel}{N}

\newcommand{\omtau}{\omega_{\tau}\z}

\newcommand{\omt}{\omega_{0} t}

\newcommand{\psit}{\Psi_t\z}
\newcommand{\psiz}{\Psi_0\z}

\newcommand{\kpsi}[1]{\ket{\Psi_{#1}\z}}

\newcommand{\nd}{n_d\z}

\newcommand{\sins}[1]{\sin^2\mleft(#1\mright)}

\newcommand{\kvac}{\ket{0}}

\newcommand{\z}{^{\phantom{\dagger}}}

\pgfplotsset{compat=1.18}
\begin{document}
\title{Real-space renormalization-group study of decoherence and revival following a sudden quench in a finite system}

\author[1]{F. D. Picoli}

\author[2]{G. Diniz }

\author*[1]{L. N. Oliveira}
\email{luizno@usp.br}

\affil[1]{\small{Instituto de Física de São Carlos, University of São Paulo, São Carlos, SP, Brazil}}

\affil[2]{\small{Charles University, Faculty of Mathematics and Physics, Prague, Czech Republic}}

\abstract{
To investigate the nonequilibrium dynamics of a quantum impurity system,  we examine two physical properties for the spinless resonant level model subject to sudden quenches of the hybridization between the impurity and the metal. We compute the time dependence of the impurity occupation and survival probability (fidelity) as probes of the dephasing induced by particle-hole excitations. For finite systems, the loss of coherence is only apparent, as discrete spectra lead to quasi-periodic dynamics and revivals when phases realign. We show that a hybrid linear–logarithmic discretization suppresses these finite-size artifacts by rendering the excitation energies incommensurate, thereby reducing revivals. Starting from the Rabi oscillations in the single-site limit of the tight-binding model describing the metal, we extend the analysis to large lattices, where damping and relaxation emerge. Our combination of analytical and numerical results offers a unified picture of the crossover from coherent oscillations to effectively irreversible decoherence.
}
\maketitle
\section{Introduction}
\label{sec:introduction}

Understanding how quantum systems exchange energy, charge, and quantum information with their environment is a long-standing central problem in physics \cite{RevModPhys.75.715,RevModPhys.76.1267,2008DHFp195317,2009GWKp115137,2010BGS657,2015LiD155135,2017NgC119.156601,2020NDC115117,2023CPA075110,2024WCB115101}. In this context, nonequilibrium studies of quantum impurity systems provide a minimal but powerful setting to investigate the mechanisms underlying the loss of coherence \cite{PhysRevB.70.235317, Kandel2021}. Beyond their fundamental interest, such studies are also essential for the understanding and control of nanoscale devices, where the coupling to external reservoirs governs transport, relaxation, and decoherence \cite{2005AnS196801,PhysRevB.98.155107,Diniz2026_,2023CPA075110,PhysRevB.92.155435}.

A paradigmatic example is the resonant-level model, the simplest realization of a localized electronic state hybridized with a continuum of conduction states \cite{PhysRevB.75.125107}. Driven out of equilibrium, by the sudden activation or deactivation of the coupling between the impurity and the environment, for instance, this noninteracting model captures the key aspects of charge transfer and energy redistribution into the bath.

A sudden change in the coupling to the bath creates a local perturbation that modifies the many-body ground state, rendering it orthogonal to the initial one in the thermodynamic limit. As a consequence, the initial state evolves into a combination of excited eigenstates of the final Hamiltonian with an infinite number of low-energy particle–hole excitations \cite{Nozieres,Nozieres71,2012MWGp235104,Picoli}, an effect known as the Anderson orthogonality catastrophe \cite{1967And1049}. In this context, the survival probability (or fidelity) of the initial state encodes how rapidly the system departs from its initial configuration due to the dephasing induced by the particle-hole excitations \cite{PhysRevLett.122.040604,m17z-4g58}.

In the thermodynamic limit, the Anderson orthogonality catastrophe gives rise to dephasing, while the hybridization generates impurity--bath entanglement. The combined effect leads to the emergence of decoherence in the quantum impurity~\cite{2013SGLp165303,m17z-4g58,PhysRevLett.122.040604,PhysRevB.50.5528,Franco2013_}, as demonstrated within the nonequilibrium Keldysh Green's function formalism \cite{PhysRevB.50.5528}.

In a finite system, even in a large one, the loss of coherence is only apparent \cite{2025Bak24466}. As the perturbation generated by the quench propagates across the lattice and reaches the boundaries, the fidelity returns close to unity. This behavior is a finite-size effect and follows from the discrete nature of the spectrum, in which the excitation energies are quantized in units of the level spacing. At sufficiently long times, the phases are approximately realigned, as they satisfy $E_n t \approx 2\pi q$ with integer $q$. Therefore, the dynamics is quasi-periodic and quasi-reversible, leading to the restoration of coherence.

This restoration arises in numerical computations of time-dependent properties and tends to affect the accuracy of the results \cite{2010BGS657}.  However, a recently proposed hybrid discretization scheme~\cite{2021FeO12254}, which combines linear and logarithmic discretizations, significantly mitigates these finite-size effects~\cite{Picoli,m17z-4g58}. In this approach, the energy scale decreases exponentially with the number of logarithmically discretized sites, so that the excitation energies are no longer commensurate. As a result, rephasing is suppressed, strongly reducing the revivals and the apparent restoration of coherence.

Here, we work with the hybrid discretization to investigate the nonequilibrium dynamics of the spinless resonant-level model following a sudden quench of the hybridization between the impurity and a metallic lattice. We focus on the time evolution of the survival probability (fidelity) and the impurity occupation.

This choice relies on the notion that, in our model, the combined decays of the fidelity and of the impurity occupation constitute a signature of decoherence. More generally, to monitor decoherence, one can track the decay of the fidelity and of the purity \cite{Franco2013_}. In the present setting, the purity is given by the expression $\mathcal{P}(t) = n_d^2(t) +(1-\nd(t))^2$, the right-hand side of which depends on the impurity occupation only. Instead of the purity, we prefer to monitor the occupation, along with the fidelity.

To build physical intuition, we start with the simplest limit, which reduces the conduction band to a single site. The exact solution of this problem exhibits Rabi oscillations reflecting the transitions between the impurity level and the conduction orbital. The persistence of these oscillations is the signature of a fully coherent evolution. We then extend the analysis to large lattices, where the spectrum of particle-hole excitations damps the analogous oscillations, leading to relaxation and effective decoherence. By contrasting the two-level and the extended systems, we provide a transparent picture of the crossover from coherent quantum oscillations to effectively irreversible decoherence.

This setting allow allows us to assess the merits of the hybrid-discretization numerical renormalization-group construction as a tool for describing the time dependence of physical properties for impurity models. To this end, we compare its performance with that of the conventional approach, which relies on a straightforward diagonalization of the model Hamiltonian, as well as with results obtained with the nonequilibrium Keldysh Green's function formalism. This comparison shows that the hybrid procedure achieves significantly higher accuracy with considerably less computational effort.

\section{Model}
\label{sec:model}
We focus our study on the \emph{resonant-level model}, the spinless version of the noninteracting Anderson impurity model \cite{Anderson1961} defined by the Hamiltonian
\begin{align}
  \label{eq:1}
  \hn{L} = \ed\cd{d} \cn{d} -\tau\sum_{j=1}^{L-1}(\cd{j}\cn{j-1}+\hc) \nonumber
   \\ +\mv(t)(\cd{d}\cn{0} +\hc),
\end{align}
where the first term on the right-hand side represents the impurity, with energy $\ed<0$, the second is the tight-binding Hamiltonian that describes the metal, and the third is the time-dependent coupling between the impurity and the metal 

The Hamiltonian $\hn{L}$ commutes with the charge operator
$\mathcal{Q}=\cd{d}\cn{d} + \sum_{j=1}^L\cd{j}\cn{j}$. 
We will let the number $L$ of lattice sites be even, and focus our attention on the half-filled system, with $\nel= \tfrac{L}2 + 1$ electrons. Section~\ref{sec:results-1} will study two time evolutions, considering a coupling that is suddenly switched off
\begin{align}
  \label{eq:2}
  \mv(t) = \goff(t) \equiv 
  \begin{cases}
    \sqrt{\tau\Gamma} & t <0\\
    0 & t \ge0\\
  \end{cases}, 
\end{align}
where $\Gamma$ is a fixed impurity-level width, and one that is suddenly switched on
\begin{align}
  \label{eq:3}
  \mv(t) = \gon(t) \equiv 
  \begin{cases}
    0 & t <0\\
    \sqrt{\tau\Gamma} & t \ge0
 \end{cases}.
\end{align}

In both cases, we will study the evolution of the quantum state $\ket{\psit}$ and calculate the
probability $\ff(t)$ that the system remains in the initial ground state:
\begin{align}
  \label{eq:4}
  \ff(t) \equiv \abs{\braket{\psit}{\psiz}}^2.
\end{align}

When we activate the coupling, the impurity level $\cn{d}$ is initially filled, that is,
$\nd(t=0)=1$, and its occupation tends to decay towards the ground-state expectation value. In the one-electron system that we discuss in the following section, $\ff(t)$ is
the probability that it returns to the initial filling. 

When we turn the coupling off, the initial occupation is smaller, $\nd(t=0)<1$, since the impurity is hybridized with the conduction band. After the quench, the
occupation remains constant. 

The model is simple, and the underlying physics is well understood. Our numerical results substantiate these notions quantitatively. However, our goal is to highlight the advantages of the discretization procedure. The comparison with the standard approach, linear discretization, makes the finite-size effects apparent, in the time dependences of both the impurity occupation and the fidelity.

\section{Single-site lattice}
\label{sec:single-lattice-site}
Before turning to long lattices, it is instructive to discuss the simplest version of our model. The choice $L=1$ eliminates the second term on the right-hand side of Eq.\eqref{eq:1}, so that the Hamiltonian becomes
\begin{align}
  \label{eq:5}
  \hn{L=0} = \ed\cd{d}\cn{d} + \mv(t)(\cd{d}\cn{0}+\hc).
\end{align}

\subsection{Coupling suddenly switched on}
\label{sec:coupling-sw-on}

Let the coupling be switched on at $t=0$ [Eq.\eqref{eq:3}]. The right-hand side of Eq.\eqref{eq:5}
is diagonal for $t<0$. Let $\kvac$ denote the vacuum state. The
two many-body eigenstates with $\nel=1$ are the ground state $\ket{\varphi_d\z} \equiv \cd{d}\kvac$, with the
energy $\epsilon_0\z= \ed$, and the excited state $\ket{\varphi_c\z}\equiv \cd{0}\kvac$, with
the energy $\epsilon_1\z=0$. At $t=0$, the quantum state is $\ket{\psiz}=\ket{\varphi_d\z}$.

  At positive times, $\hn{L}$ is non-diagonal, but its diagonalization is straightforward and
  determines the single-particle Fermi eigenoperators 
  \begin{align}
    \label{eq:6}
    a_-= \cos(\frac{\phi}2)\cn{d} +\sin(\frac{\phi}2)\cn{0}, 
  \end{align}
  and
  \begin{align}
    \label{eq:8}
    a_+ = -\sin(\frac{\phi}2)\cn{d} +\cos(\frac{\phi}2)\cn{0}, 
  \end{align}
  where
  \begin{align}
    \label{eq:10}
    \tan(\phi) = \frac{2\sqrt{\tau\Gamma}}{\ed},
  \end{align}
with the energies
  \begin{align}
    \label{eq:9}
    \omega_{\pm}\z = \omega_d\z \pm \omega_0\z, 
  \end{align}
where
  \begin{align}
    \label{eq:11}
  \omega_d\z\equiv \frac{\ed}{2}
  \end{align}
and 
\begin{align}
  \label{eq:12}
  \omega_0\z \equiv \frac{\sqrt{\eds + 4\tau \Gamma}}2.
\end{align}

Given the initial condition 
$\ket{\psiz}= \ket{\varphi_d\z}$, the standard algebra then leads to the expression
\begin{align}
  \label{eq:16}
  \kpsi{t} = &\Big(\cos(\omega_0\z t)+i\cos(\phi)\sin(\omega_0\z t)\Big) \cd{d}\kvac\ncr 
  &~~+ i\sin(\phi)\sin(\omega_0\z t)\cd{0}\kvac,
\end{align}
after gauging away the phase factor $e^{-i\omega_d\z t}$.

\subsubsection{Occupation and fidelity}
\label{sec:reduc-dens-matr}
Studies of decoherence, which typically consider spin or bosonic degrees of freedom as the environment, often focus on the off-diagonal elements of an appropriately defined reduced density matrix. Here, this procedure would be equivalent to tracing out the density matrix $\hat\rho \equiv \dm{\psit}$ over the degrees of freedom of the conduction electrons. However, for a fermionic bath, a simple check proves that this approach is unproductive, as charge conservation makes the off-diagonal components of the reduced density matrix  $\hat \rho_d(t) = n_d(t) \ket{1}\bra{1} + (1-n_d(t))  \ket{0}\bra{0}$ identically null, for arbitrary lattice sizes $L$. For this reason, we adopt the above-mentioned route and study the time dependencies of the quantum impurity occupation and of the fidelity. 

The occupation at time $t$ is the expectation value
$\nd(t)\equiv \mel{\psit}{\cd{d}\cn{d}}{\psit}$. 
From Eq.\eqref{eq:16}, we find that
\begin{align}
  \label{eq:19}
  \nd(t) = \coss{\omt}+\coss{\phi}\sins{\omt},
\end{align}
or, with the help of simple trigonometric identities,
  \begin{align}
    \label{eq:20}
    \nd(t) = 1 - \sins{\phi}\sins{\omt}.
  \end{align}

Equation\eqref{eq:16} also shows that, in the simple problem $L=0$, the occupation $\nd(t)$ is equal to the \emph{fidelity}
\begin{align}\label{eq:21}
\ff = \abs{\braket{\Psi_t\z}{\Psi_0\z}}^2
\end{align}
of the time-dependent quantum state $\kpsi{t}$ with respect to the initial state $\kpsi{0}=\ket{\varphi_d\z}$.

\begin{figure}[!t]
  \centering
  \includegraphics[width=1.0\columnwidth]{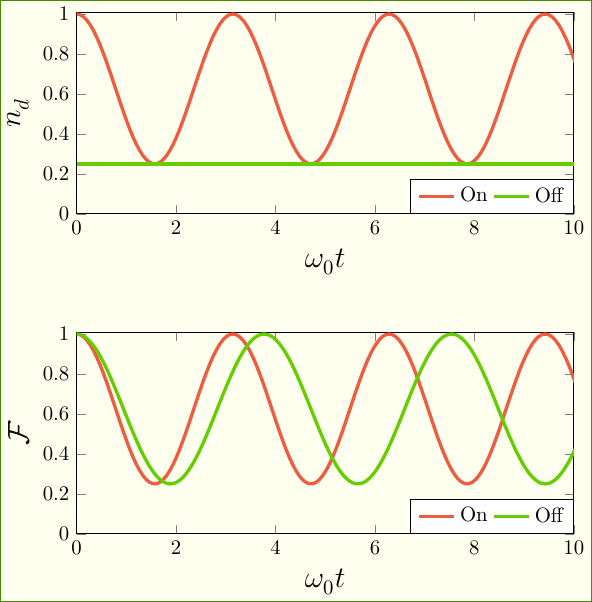}
  \caption[single-site]{Occupation of the impurity (top panel) and quantum fidelity of the state vector (bottom) as functions of time. The red curves representing the occupation and the fidelity after the coupling is turned on are identical, but the occupation remains constant after a switch-off, while the green curve representing the fidelity in the bottom panel oscillates with the frequency $\omega_d\z$ [Eq.\eqref{eq:10}].}
  \label{fig:1}
\end{figure}

As Fig.~\ref{fig:1} shows, the probability oscillates indefinitely with frequency $2\omega_0\z$ between the unitary initial value and the minimum $\cos^2(\phi)$. This Rabi oscillation is the signature of periodic transitions between the impurity and the $\cn{0}$ orbital representing the metal.

\subsubsection{Coupling switched off}
\label{sec:coupling-sw-off} Now consider an impurity that is initially coupled to the metallic orbital $\cn{0}$, and let the coupling suddenly switch off at $t=0$, following Eq.\eqref{eq:2}. The initial single-particle eigenstates are $\ket{a_\pm\z}$ [Eqs.\eqref{eq:6}~and \eqref{eq:8}] with the same eigenvalues $\omega_{\pm}$, given by Eq.~ \eqref{eq:9}.

At positive times, the eigenstates are $\ket{\varphi_d\z}$, with energy $\ed$, and $\ket{\varphi_c\z}$ with zero energy. From the initial condition $\ket{\psiz}=\ket{\varphi_-}$ results the following expression:
\begin{align}
  \label{eq:22}
  \ket{\psit}\z = \cos(\frac{\phi}2)e^{-2i\omega_d\z t}\ket{\varphi_d\z}+\sin(\frac{\phi}2) \ket{\varphi_c\z}.
\end{align}

Now, $\nd(t)$ is equal to the minimum of\eqref{eq:20}:
\begin{align}
  \label{eq:24}
    \nd(t) = \coss{\frac{\phi}2}.
\end{align}
The switch-off of $\Gamma$ at $t=0$ decouples $\cn{d}$ from $\cn{0}$. Therefore, the impurity occupation remains fixed.

In summary, whether the coupling between the impurity and the orbital representing the metal is switched on or off, the sudden change gives rise to persistent oscillations with amplitude fixed by
the ratio of energies in Eq.\eqref{eq:10}. Sections~\ref{sec:numerical-procedures}~and \ref{sec:results-1} study longer lattices and show that $\ff(t)$ is still oscillatory and so is $\nd(t)$ after the switch-on, but both the amplitude and the average value of the oscillations decay with time.

\section{Numerical procedures}
\label{sec:numerical-procedures}
At a fixed time $t$, the quadratic Hamiltonian\eqref{eq:1} can be diagonalized analytically or
numerically. If $\mv(t)=0$, so that the impurity is decoupled from the lattice, one of the
single-particle eigenvalues is $\ed$. The remaining single-particle energies are eigenvalues of the
tight-binding Hamiltonian, and can be expressed as \cite{AsM1976solid}:
\begin{align}
  \label{eq:25}
  \varepsilon_k\z = 2\tau\sin(k)\qquad k = \frac{n\pi+\frac{\pi}2}{L+1} \nonumber \\ \Big( n= -\frac{L}{2},-\frac{L}{2}+1,\ldots,\frac{L}{2}-1\Big).
\end{align}

With the coupling $\mv(t)$ on, the eigenvalues are shifted, and Eq.\eqref{eq:25} becomes
\begin{align}
  \label{eq:26}
  E_k\z = 2\tau\sin(k)\qquad k=\frac{n\pi+\frac{\pi}2-\delta_k\z}{L+1} \nonumber \\ 
  \Big(n= -\frac{L}{2},-\frac{L}{2}+1,\ldots,\frac{L}{2}-1\Big) ,
\end{align}
where the phase shifts $\delta_k$ obey the equation
\begin{align}
  \label{eq:27}
  \tan( \delta_k) = \frac{\Gamma}{\ed - E_k\z}.
\end{align}

Likewise, at fixed $t$ the single-particle eigenvectors of $\hn{t}$
can be determined analytically. However, it is more practical for our purposes to diagonalize $\hn{t}$ numerically, since a simple, very efficient
algorithm is available to compute the time dependence of the
projection $\braket{\Psi_t\z}{\Psi_0\z}$ 
from the resulting eigenvalues and eigenvectors, as described in Sec.~\ref{sec:computation-fidelity}.

\subsection{Fidelity}
\label{sec:computation-fidelity}
The algorithm we now describe is the straightforward implementation of an analytical procedure that
has been available for several decades \cite{1969NoD1097}. Let $H_I\ $ be the Hamiltonian of
interest, for either $\mv(t)=\gon(t<0)$ or $\mv(t)=\goff(t<0)$, and let $H_F\z$ be the corresponding
Hamiltonian for $t\ge0$. We denote $a_j\z$ and $\hbar\bar\omega_j$ ($j=-\tfrac{L}2,\ldots,\tfrac{L}2$)
the single particle eigenstates and eigenvalues of $H_I\z$, respectively, and $b_j$ and $\hbar\omega_j\z$
($j=-\tfrac{L}2,\ldots,\tfrac{L}2$) the single-particle eigenvalues and eigenstates of $H_F\z$. With this notation, the many-body ground state at $t=0$ is
\begin{align}
  \label{eq:28}
  \ket{\Psi_0} = \ad{0} \ad{-1}\ldots \ad{-\frac{L}{2}}\kvac,
\end{align}
and the many-body ground state at $t\ge0$ can be written as
\begin{align}
  \label{eq:29}
  \ket{\Psi_t\z} = \exp(-i\frac{H_F\z t}{\hbar})\ket{\Psi_0\z}
\end{align}
that is,
\begin{align}
  \label{eq:30}
    \ket{\Psi_t\z} = a_0^\dagger(t) a_{-1}^\dagger(t)\ldots a_{-\frac{L}{2}}^\dagger(t)\kvac,
\end{align}
where
\begin{align}
  \label{eq:31}
  a_j^\dagger(t) = \exp(-i\frac{H_F\z t}{\hbar})a_j^\dagger \exp(i\frac{H_F\z t}{\hbar}).
\end{align}

To simplify the right-hand side of Eq.\eqref{eq:31}, we express the
single-particle operator $\ad{j}$ on the basis of the eigenstates of $H_F\z$,
that is, write
\begin{align}
  \label{eq:32}
  a_j^\dagger = \sum_{k=-\frac{L}2}^{\frac{L}2} \{a_j^\dagger,b_k\z\}b_k^\dagger, 
\end{align}
and from $[H_F\z, b_k^\dagger]= \hbar\omega_k b_k^\dagger$ it follows that
\begin{align}
  \label{eq:33}
    a_j^\dagger(t) = \sum_{k=-\frac{L}2}^{\frac{L}2}
  \{a_j^\dagger,b_k\z\} e^{-i\omega_k t}b_k^\dagger,
\end{align}
and bringing Eqs.\eqref{eq:28}, \eqref{eq:30},~and \eqref{eq:33}
together we find that
\begin{align}
  \label{eq:34}
  \braket{\Psi_t\z}{\Psi_0\z} = \det{\mathcal{M}},
\end{align}
where the $(L/2+1)\times(L/2+1)$ matrix $\mathcal{M}$ comprises the
elements
\begin{align}
  \label{eq:35}
  \mathcal{M}_{j,k} =
  \sum_{\ell=-\frac{L}2}^{\frac{L}2}\{a_j,b^\dagger_\ell\} e^{-i\omega_\ell\z t}\{b_\ell,a_k^\dagger\}.
\end{align}

Therefore, the computation of the projection
$\braket{\Psi_t\z}{\Psi_0\z}$ requires the calculation of a single
determinant at each time. In practice, a run that computes the
projection at $1\,000$ times ranging from $t=0$ to $t=2\,000$ for a
lattice with $L=2\,000$ sites takes roughly \SI{100}{sec} on a
standard desktop computer.

\subsection{Impurity occupation}
\label{sec:comp-impur-occup}
If the coupling is switched off, the impurity charge is conserved and can be readily
computed from the expression
\begin{align}
  \label{eq:36}
  \nd(t=0) = \sum_{p<p_F\z}\{\cd{d},\bn{p}\} \{\cn{d}, \bd{p}\},
\end{align}
where the sums are restricted to eigenstates below the Fermi level. Alternatively, the equilibrium occupation can be obtained from the Fermi level phase shift, given by Eq.\eqref{eq:27}, with the help of the Friedel sum rule.

As explained in the introduction, our criterion for decoherence in the resonant-level model is the decay of both $\ff(t)$ and $\nd(t)$. The time independence of the latter after the switch-off is aligned with this criterion, as the deactivation of the coupling dissociates the impurity from the time evolution of the metal. Under these circumstances, the fidelity monitors only the dephasing within the conduction band, a restricted form of decoherence.

In contrast, when the coupling is suddenly switched on, the impurity filling is unitary before
$t=0$, but becomes time dependent after that. The expectation value $\nd(t)$ is then given by the
expression
\begin{align}
  \label{eq:37}
  \nd(t) = \mel{\psit}{\cd{d}\cn{d}}{\psit}, 
\end{align}
equivalent to 
\begin{align}
  \label{eq:38}
  \nd(t) = \mel{\psiz} {e^{iH_F\z t}\cd{d}\cn{d}e^{-iH_F\z t}}{\psiz},
\end{align}
where $\kpsi{0}$ is the ground-state of the initially uncoupled Hamiltonian. 

The impurity number operator $\cd{d}\cn{d}$ commutes with the initial Hamiltonian, but not with
the final one. It is therefore appropriate to expand $\cn{d}$ on the single-particle basis of
the eigenstates of $H_F\z$:
\begin{align}
  \label{eq:39}
  \cn{d} = \sum_{p} \{\cn{d}, \bd{p}\} \bn{p}
\end{align}
and rewrite Eq.\eqref{eq:38} as
\begin{align}
  \label{eq:40}
  \nd(t) = \sum_{p,q}\{\cd{d},\bn{p}\}\{\cn{d},\bd{q}\}e^{i(\omega_q\z-\omega_p\z) t}
  \mel{\psiz}{\bd{p}\bn{q}}{\psiz}.
\end{align}

The evaluation of the matrix element on the right-hand side of Eq.\eqref{eq:40} is straightforward and shows that
\begin{align}
  \label{eq:41}
  \mel{\psiz} {\bd{p}\bn{q}}{\psiz} =\sum_{m<m_F\z} \{\an{m},\bd{p}\}\{\ad{m},\bn{q}\},
\end{align}
where the sum is restricted to the single-particle eigenstates $\an{m}$ below the Fermi level.

The combination of Eqs.\eqref{eq:40}~and \eqref{eq:41} then produces the result
\begin{align}
  \label{eq:42}
  \nd(t) = \sum_{p,q}\{\cd{d},\bn{p}\}\{\cn{d},\bd{q}\}e^{i(\omega_q\z-\omega_p\z) t} \times \nonumber \\
  \sum_{m<m_F\z}\{\an{m},\bd{p}\}\{\ad{m},\bn{q}\}.
\end{align}

At $t=0$, the exponential on the right-hand side being unitary, the sums over $p$ and $q$ can be
eliminated, and $\nd$ becomes equal to unity, since the eigenstate
$\ket{\varphi_d\z}$ of the initial Hamiltonian lies below the Fermi level and is fully included in the remaining sum. However, at long times only final-state eigenstates with low energy differences can contribute, and the right-hand side
of Eq.\eqref{eq:42} is expected to approach that of Eq.\eqref{eq:36}.

\subsection{eNRG method}
\label{sec:enrg-method}
As discussed in Sec.~\ref{sec:computation-fidelity}, the evaluation of the fidelity and impurity occupation for
our model is a relatively simple numerical task. However, the resonant-level
Hamiltonian~\eqref{eq:1} exhibits a feature that is detrimental to the computation of
time-dependent properties. At long times, such that $t\gg \max\left\{\tau, \tfrac{1}{\Gamma}, \tfrac{1}{\abs{\ed}}\right\}$, the low energy spectrum of this Hamiltonian controls the dynamics of the model. Under this condition, the linearization of Eq.\eqref{eq:25} gives us the approximate
expression
\begin{align}
  \label{eq:43}
  \varepsilon_k\z = \Big(n+\frac12\Big)\Delta\varepsilon\qquad(n=0, \pm1,\ldots),
\end{align}
where
\begin{align}
  \label{eq:44}
  \Delta\varepsilon =\frac{2\pi\tau}{L}.
\end{align}

We can see that the many-body energies $E$ resulting from combinations of the single-particle energies\eqref{eq:43} are multiples of $\Delta\varepsilon$. Therefore, at time $t= T_{2\pi}\z\equiv2\pi/(\Delta\varepsilon)$, all the factors $e^{-i E t}$ in the time evolution of the physical properties return to their value at $t=0$. The computation of
time-dependent properties produces artificially periodic results, with periodicity
$T_{2\pi}\z=\frac{L}{\tau}$. In practice, this finite-size effect introduces significant
deviations well before  $t=T_{2\pi}\z$, as the numerical results in
Sect.~\ref{sec:coher-tight-bind} show. We call this limitation the \emph{commensurability
  problem} and turn to an alternative description of the conduction band
that avoids commensurability, the eNRG approach \cite{2022FeO075129}.

The eNRG method is a real-space version of the construction of the numerical renormalization group (NRG)
\cite{Wilson_NRG,Krishna_murthy_1,Bulla}. The resulting procedure is more flexible
because it involves three parameters, instead of two: the \emph{discretization} parameter $\lambda$,
an integer larger than unity, analogous to the NRG parameter $\Lambda$, the \emph{twist} $\theta$,
analogous to the NRG parameter $z$, and the \emph{offset} $\zeta$, a positive integer with no
parallel in the NRG construction. Here, for simplicity, we will work with a fixed $\theta=0$.

The eNRG construction is applied to the lattice defined in Sec.~\ref{sec:model}, with $L$ sites and
the $L$ Fermi operators $\cn{\ell}$ ($\ell=0,1,\ldots, L-1$). Our first goal is to define an equivalent
basis with a substantially smaller number $\mn+1 \ll L$ of Fermi
operators. We let the first $\zeta$ normalized Fermi operators $a_j\z$ coincide with those
in the original basis: $a_{m}\z = c_{m}\z$ ($m=0,1,\ldots, \zeta-1$). Starting from
$\ell=\zeta$, we group the sites into cells $\mc{m}$ ($m=0,1,\ldots$) containing $\lambda^m$ sites.

For each cell $\mc{n}$ we lump the $\lambda^n$ operators $\cn{\ell}$ into a single operator of the
new basis,
\begin{align}
  \label{eq:45}
  \fn{n} = \alpha_n\z \sum_{\ell\in\mc{n}}e^{i\frac{\pi}{2}\ell}\cn{\ell},
\end{align}
where the factor $\alpha_n\z = \lambda^{-\frac{n}2}$ normalizes $\fn{n}$.

The projection of the Hamiltonian $\hn{L}$ on the basis of the operators $\cn{d}$, $\an{n}$
($n=0,1,\ldots,\zeta-1)$ and $\fn{m}$ ($m=0,1,\ldots$) results in the eNRG Hamiltonian
\begin{align}
  \label{eq:46}
  \hnrg =& \ed\cd{d} \cn{d}   +\mv(t)(\cd{d}\an{0} +\hc) \nonumber \\
           &-\tau\Big(\sum_{j=1}^{\zeta}\ad{j}\an{j-1}+ \hc\Big)\ncr
           &+\tau \big(\ad{\zeta-1}\fn{0}  +\hc\big) \nonumber \\
  &+\tau \Big(\sum_{m=0 }^{M-1}\lambda^{-{n-1}/2}\fd{m}\fn{m+1}+\hc\Big).
\end{align}

If we were to drop the last term on the right-hand side, the substitution $\zeta\to L$ would reduce
 Eq.\eqref{eq:46} to $\hn{L}$. The eNRG lattice is equivalent to the
tight-binding lattice augmented by $M$ sites with progressively decaying couplings. The algorithm
that computed the fidelity in Sec.~\ref{sec:computation-fidelity} can be used to calculate the
projection from the Hamiltonian\eqref{eq:46} as well. The costs of computing $\ff(t)$ from $\hn{L}$ and $\hnrg$ are similar, but the
latter produces significantly more precise results, as Sect.~\ref{sec:results-1} shows.

\section{Results}
\label{sec:results-1}
\begin{figure}
    \centering
\includegraphics[width=1.0\linewidth]{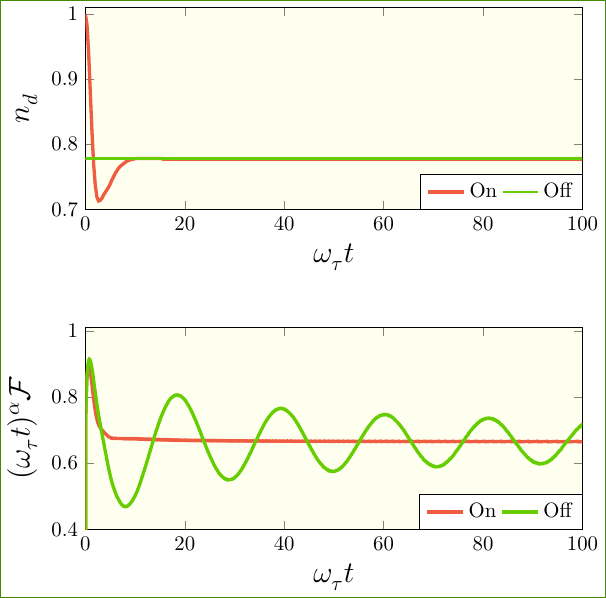}
    \caption{Impurity occupation $\nd$ and quantum fidelity $\ff$ as functions of time after the
      coupling to the conduction band is switched on (red) or off (green curves). All curves were
      computed with impurity energy $\ed=-0.3\,\tau$ and width $\Gamma=0.3\,\tau$. The fidelities
      have been scaled by the Doniach-Sunjic power law with an exponent $\alpha=2(\tfrac{\delta}\pi)^2$, where $\delta$ is the conduction-electron phase shift at the
      Fermi level. When the coupling is switched off,  the occupation remains constant after the
      quench, but the fidelity oscillates with decaying amplitude. By contrast, when the coupling is
      switched on, the occupation rapidly drops from unity and approaches the straight line representing
      the occupation after the coupling is switched off, while the scaled fidelity rapidly approaches the average value of the green curve representing the fidelity after the coupling is switched off.}
    \label{fig:2}
\end{figure}

Figure~\ref{fig:2} surveys our findings. The displayed curves were computed from the eigenvalues and eigenvectors resulting from the numerical diagonalization of the eNRG Hamiltonian\eqref{eq:46} with
$\lambda=1.5$, $\zeta= 100$, and $M=20$, substituted in Eq.\eqref{eq:34}, for the fidelity, and
Eqs.\eqref{eq:36}~and \eqref{eq:42} for the impurity occupation. The red and green solid lines describe suddenly switched-on or off couplings, respectively. As expected, the time dependence of the impurity
occupation depends on whether the coupling is turned on or off. Specifically, under a switch-off, the decoupling fixes the
occupation. Under a switch-on for the same model parameters, $\nd$ is a function of time that decays from $\nd(0)=1$ and approaches the same constant as $t\to\infty$. However, the decay from $\nd(0)=1$ to $\nd(t\to\infty)$ is not monotonic. As the red plot shows, the occupation overshoots the final value before rising to stability.

The contrast between the fidelities after a switch-on and a switch-off is more striking. In both cases, the
fidelity ultimately decays in conformity with the Doniach-Sunjic law. However, the decay is monotonic after the coupling is turned on but displays weakly damped oscillations of frequency $\abs{\ed}$ after a switch-off. This behavior is reminiscent of the Rabi oscillations of the green curve in the bottom panel of Fig.~\ref{fig:1} and has the same physical origin: the interference between many-body states in which the impurity is occupied and states in which it is vacant.

A more detailed discussion of the physics seems appropriate. Consider the turn-off first. In the initial ground state $\ket{\Omega_0}$, the impurity is entangled to the conduction bath in a coherent superposition of occupied ($\nd=1$) and vacant ($\nd=0$) states. After the coupling is switched off, the conservation of $\nd$ divides the spectrum into two sectors, with $\nd=1$ and $\nd=0$. We will refer to the many-body states in the former (latter) sector as the \emph{occupied} (\emph{vacant}) states. The quantum state $\ket{\psit}$ is now a  linear combination of a vacant quantum state $\ket{\Psi^0_t)}$ and an occupied quantum state $\ket{\Psi^1_t}$. Neither $\ket{\Psi^0_t}$ nor $\ket{\Psi^1_t}$ is an eigenstate of the model Hamiltonian. 
Both are linear combinations of numerous particle-hole excitations. 

The energy difference between an occupied state and the corresponding vacant  state is $\ed$. Therefore, relative to  the $\nd=0$ quantum state, the $\nd=1$ state carries the phase factor $\exp(-i\ed t)$. The oscillations of the green curve in the bottom panel of Fig.~\ref{fig:2} express the interference between $\ket{\Psi_t^0}$ and $\ket{\Psi_t^1}$.

The power-law decay of the same curve is due to the Anderson catastrophe \cite{1967And1049,2023MISp186401,PhysRevLett.122.040604,2014GKM,2013SGLp165303}. The projection between the conduction bands in the initial and final quantum states determines its exponent. When the initial and final conduction bands accommodate the same number of electrons, as in X-ray photoemission spectroscopy, at long times, the fidelity follows the power law
\begin{align}
  \label{eq:47}
  \ff(t) \sim (\omtau t)^{-\alpha},
\end{align}
with the shorthand $\omtau \equiv \tfrac1\tau$ and the Doniach-Sunjic exponent \cite{1970DoS285}
\begin{align}
  \label{eq:48}
  \alpha = 2\left(\frac{\delta}{\pi}\right)^2,
\end{align}
where $\delta$ is the difference between the phase shifts $\delta_F$ and $\delta_I$ near the Fermi levels of the final and initial conduction bands, respectively. 

When the final conduction band contains one more or one less electron than the initial band, as in X-ray absorption spectroscopy, the power law is
\begin{align}
  \label{eq:49}
  \ff(t) \sim (\omtau t)^{-\beta},
\end{align}
with the Nozières-De Dominicis exponent \cite{1969NoD1097}
\begin{align}
  \label{eq:500}
  \beta = 2\left(1-\frac{\delta}{\pi}\right)^2.
\end{align}

After the coupling $\mathcal{V}$ is switched off, the $\nd=0$ state $\ket{\Psi_t^0}$ contains $\tfrac{L}2+1$ electrons, the same number as in $\ket{\Omega_0}$. Therefore, the contribution of $\ket{\Psi_t^0}$ to the fidelity follows the Doniach-Sunjic power law\eqref{eq:48}. 
In contrast, the $\nd=1$ state $\ket{\Psi_t^1}$ contains $\tfrac{L}2$ conduction electrons. The 
contribution of $\ket{\Psi_t^1}$ to the fidelity decays with the Nozières-De Dominicis law~\eqref{eq:500}. The Fermi-level phase shift $\delta_I$ before the switch-off is given by the $E_k\to 0$ limit of Eq.\eqref{eq:27}:
\begin{align}
  \label{eq:51}
  \tan(\delta_I) = \frac{\Gamma}{\ed},
\end{align}
while the phase shift after the quench is $\delta_F=0$.

With the parameters in Fig.~\ref{fig:2}, the initial phase shift is $\delta_I=-\tfrac{\pi}4$, and the two exponents are $\alpha = \tfrac18$ and $\beta= \tfrac98$. Therefore, the contribution from the occupied state $\ket{\Psi_t^1}$ decays faster than the contribution from $\ket{\Psi_t^0}$, progressively reducing the interference, as illustrated by the damped oscillations of the green curve in the bottom panel of Fig.~\ref{fig:1}.

Next, consider the switch-on. The initial ground state $\ket{\Omega_0}$ comprises the impurity and conduction band, decoupled from each other. The impurity is occupied, and the band is half-filled. The Fermi-level phase shift is $\delta_I=0$.
The activation of the coupling $\mathcal{V}$ starts the evolution of the quantum state $\ket{\psit}$ from its initial condition $\ket{\psiz} =\ket{\Omega_0}$.

The electronic transitions between the impurity and the band allows the transfer of charge across the Fermi level, creating particle-hole excitations that drive the system towards equilibrium.  The equilibrium impurity density of states, which controls the dynamics of this process, peaks at $\ed$ and has the width $\Gamma$. Therefore, the characteristic time of the relaxation towards equilibrium is $\tfrac1{\Gamma}$, and time-dependent physical properties must oscillate with frequency $\abs{\ed}$ and damping $\Gamma$. 

The fidelity is no exception. At any instant $t$, the state $\ket{\psit}$ is a linear combination of occupied and vacant states. The energy differences between these $\nd=1$ and $\nd=0$ states form a distribution of width $\Gamma$ peaked at $\ed$. The interference between the contributions of the occupied and vacant states gives rise to fidelity oscillations that decay on the time scale of $\tfrac{1}{\Gamma}$.

The Anderson catastrophe introduces an additional source of damping. In the initial state $\ket{\Omega_0}$, the conduction band contains $\tfrac{L}2$ electrons. After the quench, the occupied and vacant states contain $\tfrac{L}2$ and $\tfrac{L}2+1$ conduction electrons, respectively. Therefore, the evolution of the occupied states is governed by the Doniach-Sunjic power law, while the evolution of vacant states follows the Nozières-De Dominicis law. Under the conditions of Fig.~\ref{fig:2}, the initial and final phase shifts are $\delta_I=0$ and $\delta_F= -\tfrac{\pi}4$, and the exponents are $\alpha=\tfrac{1}{8}$ and $\beta= \tfrac{25}8$. 
Therefore, the contribution of the vacant states to $\ff(t)$ vanishes a great deal faster than the contribution from the occupied states, reducing the effects of interference and damping the oscillations.

The damping due to the transitions between the impurity and the band  reduces the amplitude of the oscillations exponentially, with the characteristic time $1/\Gamma$.  The decay due to the orthogonality catastrophe is devoid of characteristic times and follows a power law. Therefore, the former source of damping is dominant. 

With $\Gamma = 0.3\tau$, this source is so strong that the undulation of the red curve representing the fidelity under a switch-on in the bottom panel of Fig.~\ref{fig:2} is barely perceptible.

As $\Gamma$ is reduced, the characteristic time grows. while the Nozières-De Dominicis exponent $\beta$ approaches unity. Therefore, both sources of damping are reduced. The undulations in the plots of $\ff(t)$ then become clearly visible, as the following section shows.

\subsection{Fidelity}
\label{sec:suddenly-off}
\begin{figure}[!ht]
  \centering
 \includegraphics[width=1.0\linewidth]{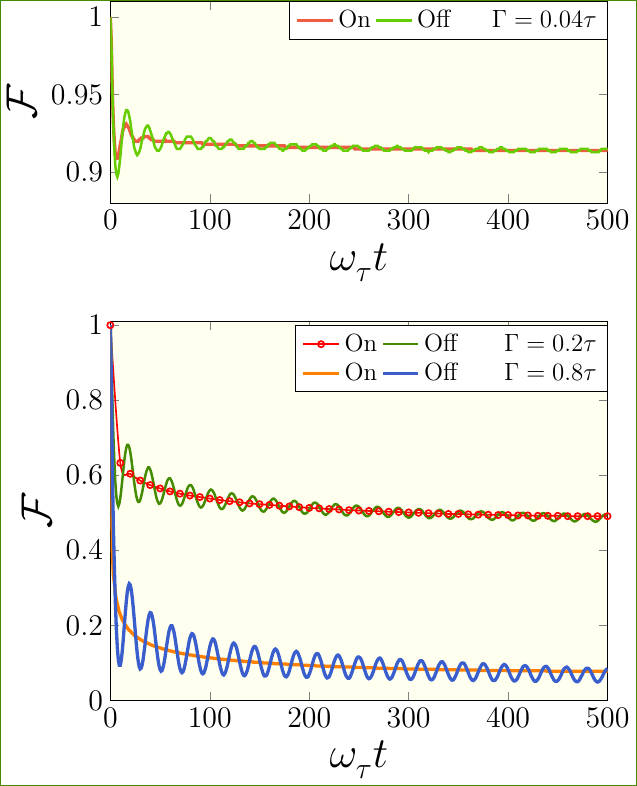}
 \caption[Fid]{Decay of the quantum fidelity after the coupling between the impurity and the metal is suddenly turned on or off, for the indicated hybridizations $\Gamma$. For each $\Gamma$ the
   behaviors after the sudden switch-on and switch-off are shown. All decays
   follow the Doniach-Sunjic law, the exponent $\alpha$ [Eq.\eqref{eq:48}] increasing with
   $\Gamma$. When the coupling to the impurity is deactivated, the contributions to the fidelity from the $\nd=1$ states, which are governed by the Nozières-De Dominicis exponent $\beta$ [Eq\eqref{eq:49}], decay faster, progressively reducing the interference and damping the oscillations. When the coupling is activated, the contributions to the fidelity from electronic transitions from the impurity to various particle or hole energies in the conduction band damps the oscillations on the time scale of $\tfrac{1}{\Gamma}$.}
 \label{fig:3}
\end{figure}

After discussing the main features of the response to a sudden quench, we now present the results for the quantum fidelity $\ff(t)$ more systematically. Figure~\ref{fig:3} displays the eNRG results for $\ed=-0.3\tau$ and three representative $\Gamma$, under switching on and
off. All results were obtained with $\lambda=1.5$, $\zeta=300$, and $M=16$. 

 The top panel shows the decays for a small coupling, with $\Gamma=0.04\tau$. Since $\Gamma\ll |\ed|$, this coupling introduces a minute phase shift at the Fermi level,
 given by Eq.\eqref{eq:27}. The Doniach-Sunjic exponent being close to zero, the decay is slow. The Nozières-De Dominicis exponent $\beta$ is close to unity, both for the switch-on and off. When the coupling is deactivated, the decay of the contributions from the occupied states damps the oscillations of the solid green curve. 
 
 When the coupling is activated, the dominant source of damping is the transfer of charge between the impurity and the conduction band, which eliminates the undulations on the time scale of $1/\Gamma$. Accordingly, the red curve in the top ($\Gamma=0.04\tau$) panel of Fig.~\ref{fig:4} displays a few cycles of oscillation, while the red circles and solid line in the bottom panel ($\Gamma=0.2\tau$ and $\Gamma=0.8\tau$, respectively) decay monotonically.

On intuitive grounds, one would expect the loss of coherence to be mirrored in the time dependence of the impurity occupation. We have already seen evidence of this in Fig.~\ref{fig:2}, and the data in the following section reinforce this notion.

\subsection{Impurity occupation}
\label{sec:suddenly-on}
\begin{figure}[!ht]
  \centering
    \includegraphics[width=1.0\linewidth]{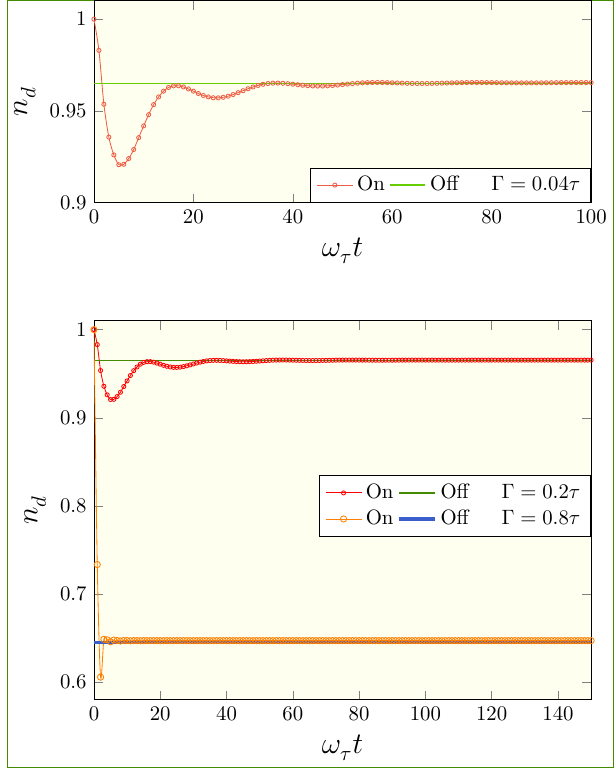}
  \caption[nd]{Impurity occupation following a sudden quench, for the
    impurity energy $\ed=-0.3\tau$ and the indicated hybridizations $\Gamma$. When the coupling is switched
    off, the impurity occupation retains the initial ground-state expectation value $n_d^0 = \mel{\Psi_0\z}{n_d\z}{\Psi_0\z}$ and therefore remains constant, as illustrated by the green and the blue
    horizontal lines in the two panels. Upon switch-on, the occupation oscillates with frequency $\abs{\ed}$, but the oscillations are progressively more damped as $\Gamma$ increases.}
  \label{fig:4}
\end{figure}

The curves in Fig.~\ref{fig:4} are in strict correspondence to the fidelity plots in
Fig.~\ref{fig:3} and were extracted from the same eNRG runs. Of central interest are the behaviors
after the coupling is switched on. For each of the three illustrative widths $\Gamma$, the occupation decays
from the initial value $\nd(0)=1$ and approaches the off-switch occupation $n_d^0$, which is represented
by the horizontal straight line. The decay is a damped oscillation, the damping rising rapidly with
$\Gamma$, in analogy with the fidelity curves under the on-switch in Fig.~\ref{fig:3}. As discussed in Sec.~\ref{sec:suddenly-off}, the activation of the coupling to the impurity leads to readjustment of the electronic density. This accommodation, aided by the Anderson catastrophe, damps the oscillations and drives the impurity density to its equilibrium value.

\subsection{Coherence in the tight-binding model}
\label{sec:coher-tight-bind}
Secs.~\ref{sec:suddenly-off}~and \ref{sec:suddenly-on} have focused on the physical aspects of the results. The eNRG data in Secs.~\ref{sec:suddenly-off}~and \ref{sec:suddenly-on} have shown that, following a switch-off, the Anderson catastrophe, through the Doniach-Sunjic and Nozières-De Dominicis mechanisms, progressively dephase the Rabi oscillations created by the quench. Following a switch-on, the readjustment of the electronic distribution is the dominant source of decoherence.

We now come to the central purpose of our work and discuss the virtues of the eNRG construction. To bring to light the consequences of the commensurability problem discussed in Sec.~\ref{sec:enrg-method}, we compare the time dependences obtained directly from the tight-binding Hamiltonian~\eqref{eq:1} with those obtained from the eNRG Hamiltonian~\eqref{eq:46}.

\subsubsection{Fidelity}\label{sec:sub.fidelity}

\begin{figure}[!ht]
  \centering
\includegraphics[width=1.0\linewidth]{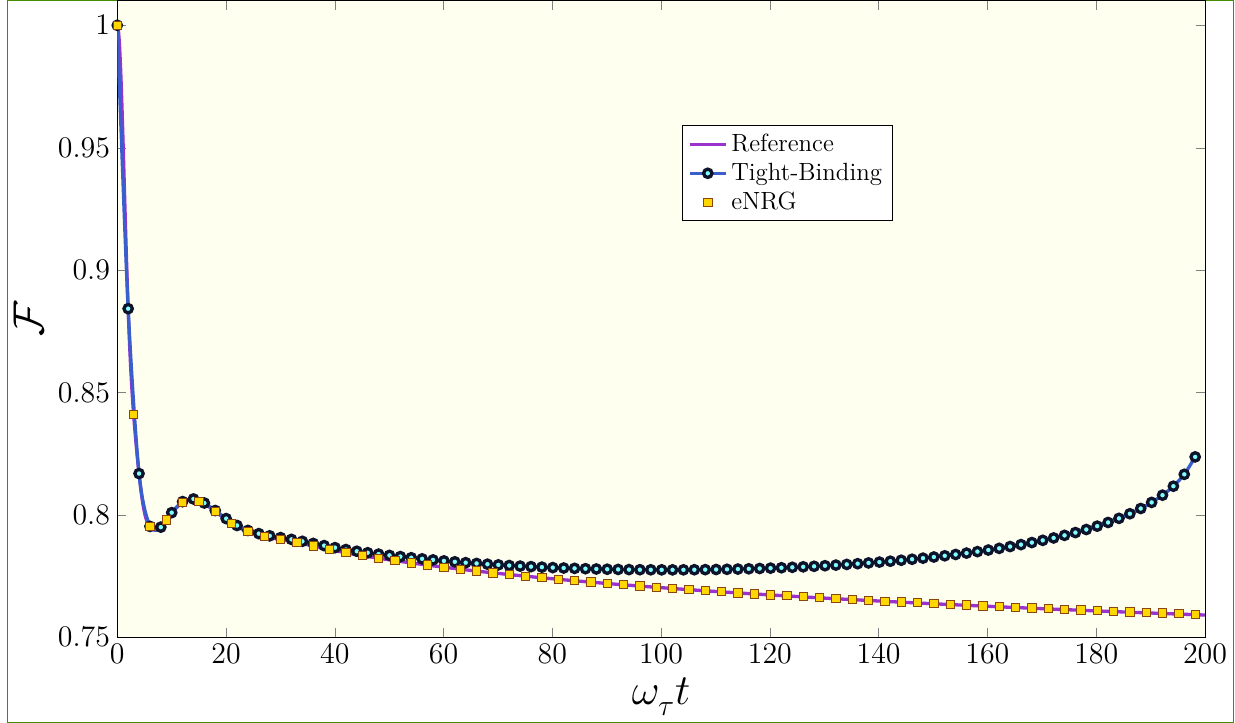}
  \caption[tbind]{Fidelity as a function of time for a tight-binding lattice coupled to an impurity with energy
  $\ed=-0.3\tau$ and hybridization $\Gamma=0.1\tau$. The coupling
  between the metal and impurity is turned on at $t=0$. The purple
  curve, here adopted as a benchmark, represents the essentially exact fidelity computed by the eNRG large $\zeta$ procedure, with $\lambda=1.5$, $\zeta= 980$ and  $M=16$. The gold squares, in excellent agreement with this reference curve, are the result of an eNRG computation with $\lambda=1.5$, $\zeta=190$, and
 $M=10$. The dark green circles resulted from the direct diagonalization of the model Hamiltonian, Eq.\eqref{eq:1}, with $L=200$.}
  \label{fig:5}
\end{figure}

 Figure~\ref{fig:5} compares a very accurate computation of the fidelity with (i) results
obtained from directly diagonalizing the  Hamiltonian\eqref{eq:1} and (ii) eNRG data.
In both cases, the fidelities were calculated from Eq.\eqref{eq:34}.

The solid purple curve reports essentially exact results for the quantum fidelity in the displayed time
range, obtained from an eNRG computation with $\lambda=1.5$, $\zeta=980$ and $M=20$.
Exhaustive tests have shown that these results are virtually identical to data 
obtained with larger $\zeta$s and various $\lambda$ within the time interval covered by the
plot. We adopt them as a benchmark. The green circles resulted from the numerical diagonalization of $\hn{L=200}$, and the gold squares describe the eNRG data obtained with $\lambda=1.5$, $\zeta=190$, and $M=16$. The latter curve is indistinguishable from the reference curve at all times. In contrast, the green circles, which follow the red line reasonably well until roughly one-half the maximum time, rise away from the benchmark at the end of the interval. This deviation, a well-known consequence of commensurability, brings back part of the coherence that the Anderson catastrophe should have destroyed. The eNRG construction eliminates commensurability to produce fidelity curves with no trace of this artificial revival.

The impact of commensurability upon the computation of $\nd(t)$ is less pronounced. However, this time dependence deserves discussion, as it reveals another facet of the problem.  

\subsubsection{Impurity occupation}\label{sec:sub.occupation}
In contrast to the fidelity, the impurity occupation $n_d(t)$ can be computed analytically, within the Keldysh formalism.
The resulting expression, valid for small $\abs{\epsilon_d}$ and $\Gamma$, reads \cite{2006AnS245113}
\begin{align}
  \label{eq:50}
  n_d(t) = &1 - \frac{\Gamma}{\pi} \times \nonumber \\ &\int_{-\tau}^{0}\frac{1+e^{-2\Gamma t}-2e^{-\Gamma t}\cos((\epsilon-\ed)t)}{(\epsilon-\ed)^2+\Gamma^2}
  \,\dd\epsilon.
\end{align}

It would be straightforward to evaluate the integral on the right-hand side, but the clearer physical interpretation afforded by Eq.\eqref{eq:50} makes it more appealing. The numerator of the integrand is the law of cosines describing the sum of two quantum vectors separated by the time-dependent phase $(\epsilon-\epsilon_d)t$. One of the vectors has unit magnitude, while the other has the magnitude $\exp(-\Gamma t)$, which decays as the transitions between the impurity level and the conduction band drive the electronic distribution towards the equilibrium density given by the Lorentzian of width $\Gamma$ centered at $\ed$. The same physics has emerged from a recently derived expression for the time dependence of the current in X-ray photoemission \cite{Picoli}, with two important distinctions: (i) the Anderson catastrophe, instead of the electronic transitions between the impurity and the band, governs the decay of the second quantum vector, and (ii) at long times the current approaches zero, instead of the equilibrium impurity occupation.

The Keldysh treatment underlying Eq.\eqref{eq:50} is based on non-equilibrium Green's function and describes single electron-hole excitations accurately. However, as evident from Eq.\eqref{eq:50}, this formalism leaves aside the consequences of the orthogonality between the initial ($t<0$) and final ($t>0)$ ground states. 

As the fidelity plots in Figs.~\ref{fig:2}, \ref{fig:3},~and \ref{fig:5} indicate, the eNRG computations describe these effects accurately. Considering this, we have extended the range of the eNRG computations to investigate the impact of the orthogonality catastrophe on the time dependence of the impurity occupation. A comparison with Eq.\eqref{eq:50} then showed that the orthogonality enhances the damping of the oscillations. However, this enhancement is minute, so small that it cannot be seen in the following figures. Only
at times $t>\tfrac{1}{\Gamma}$, with $\Gamma\ll \tau$, does it become significant. The fidelity plots in Figs.~\ref{fig:2}, \ref{fig:3}, and \ref{fig:5} illustrate the consequences of the orthogonality catastrophe much more clearly.

\begin{figure}
\includegraphics[width = 0.95\columnwidth]{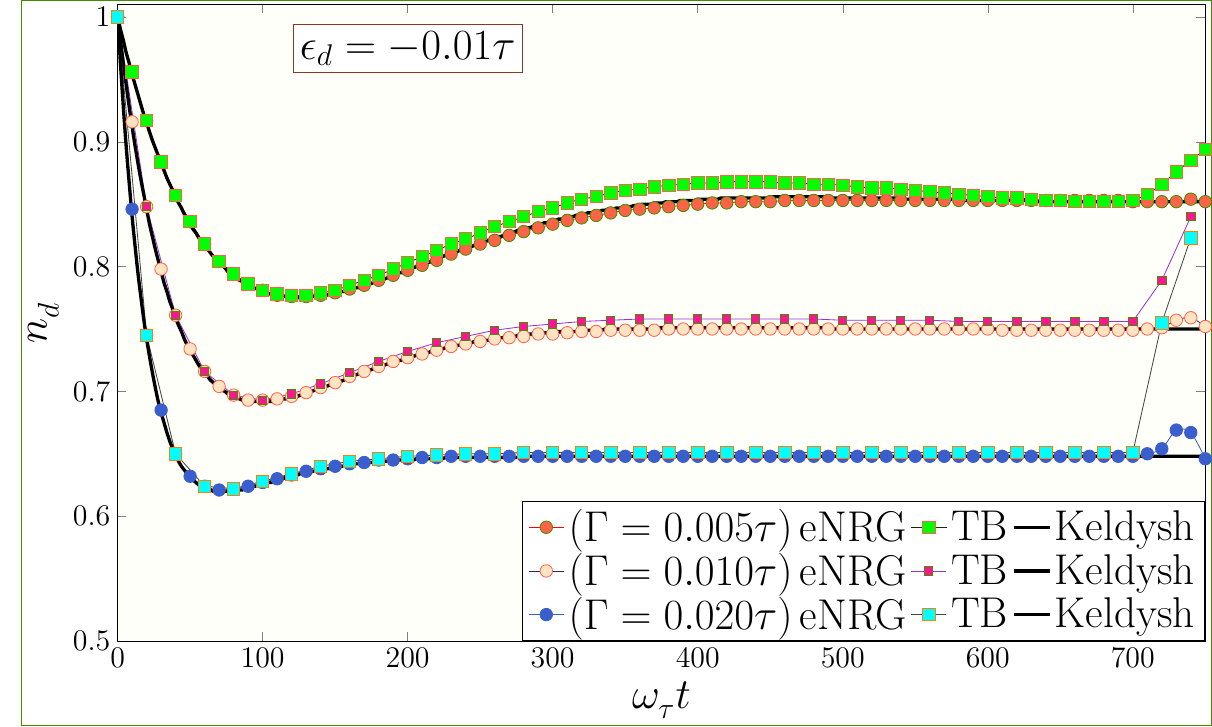}
\caption{\label{fig:6}Time dependence of the impurity occupation for $\ed=-0.01\tau$ and three level widths, $\Gamma = 0.005\tau$, $0.01\tau$, and $0.02\tau$. For each width, the occupations $n_d(t)$ numerically calculated from Eq.\eqref{eq:46} for the Hamiltonians~\eqref{eq:1}~and \eqref{eq:42} are shown as open circles, labeled \emph{eNRG}, and open squares, labeled $\emph{TB}$, respectively, compared with the solid lines representing the right-hand side of Eq.\eqref{eq:50}.}
\end{figure}

Figure~\ref{fig:6} shows $\nd(t)$ for $\ed=-0.01\tau$ and the indicated impurity-level widths $\Gamma$. For each width, open circles and squares represent the occupations for the model Hamiltonian\eqref{eq:1} and the eNRG Hamiltonian\eqref{eq:42}, respectively; the solid lines represent Eq.\eqref{eq:50}.

All circles were computed with the eNRG parameters $\lambda = \sqrt{3}$, $\zeta=700$, and $M=16$. The offset defines the interval $\omtau t <\zeta$ within which the eNRG computation produces reliable results. Physically, when the coupling to the impurity is activated, the switch-on immediately affects the orbital $\an{0}$, which is directly coupled to $\cn{d}$. The perturbation then travels along the lattice and reaches the site $c_{j=\zeta}$ at the time $\zeta/\omtau$. After that, the perturbation leaves the offset zone and encounters the exponentially decaying couplings $\tau\lambda^{-n-\tfrac12}$. The computed impurity densities then oscillate on the logarithmic scale. This well-established signature of NRG calculations \cite{z_trick_Oliveira}
is clearly visible in the $\omtau t> \zeta=700$ region of Fig.\eqref{fig:6}.

For comparison, the squares in the three plots were computed with the lattice size $L=700$. As described above, the perturbation generated by the switch-on advances and reaches $c_{j=L}$ at the time $L/\omtau$. Beyond that, the commensurability problem reverses the decay and strongly pushes the $\nd(t)$ curves upward.

Within the $\omtau<\zeta$ interval, the three sets of circles are in very good agreement with the solid black curves. The agreement between the cyan squares and the lower solid line is also very good, but the inspection of the $\Gamma=0.01\tau$ curve shows a small deviation separating the red squares from the black line. The separation grows as the width $\Gamma$ is reduced, as indicated by the green squares forming the top curve ($\Gamma=0.05\tau$).

\begin{figure}
    \centering
\includegraphics[width=0.95\linewidth]{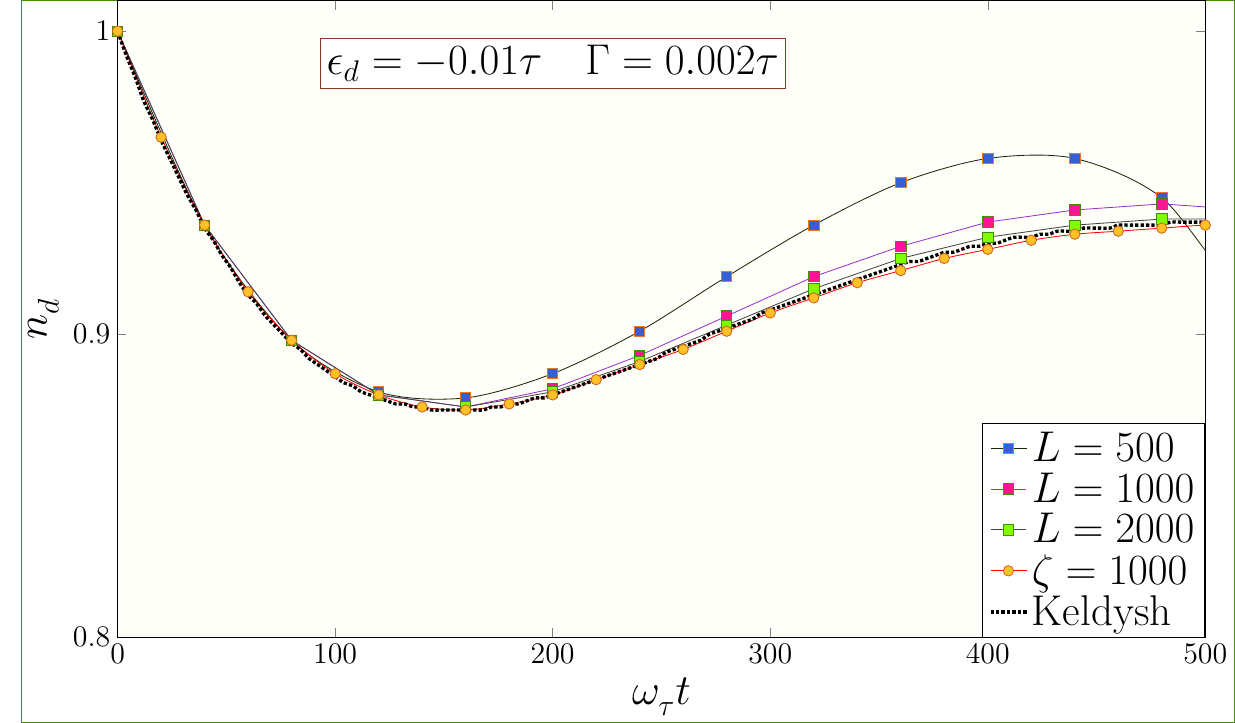}
    \caption{Impurity occupation as a function of time, computed from Eq.\eqref{eq:42} for the tight-binding Hamiltonian\eqref{eq:1} with the displayed impurity energy $\ed$ and level width $\Gamma$, compared with the exact expression\eqref{eq:50}. The blue, red, and green squares represent the occupations computed for lattices with $L=500$, $L=1000$, and $L=2000$ sites, respectively. The golden circles represent the eNRG results for the occupation. The dotted black line represents Eq.\eqref{eq:45}.}
    
    \label{fig:7}
\end{figure}

Figure~\ref{fig:7} is focused on the lattice-size dependence of these deviations for the small width $\Gamma=0.005\tau$. The separation between the squares and the dotted line representing Eq.\eqref{eq:50} shrinks as the system size grows. This evolution with $L$ reveals another facet of the commensurability problem. When  the conduction-band level separation\eqref{eq:34} becomes comparable to the width $\Gamma$, the tight-binding model leaves the $\Gamma \gg \Delta\epsilon$ regime that is characteristic of the continuum. The calculated physical properties then become functionals of $L$, an artifact of Eq.\eqref{eq:1} that is minimized by the eNRG approach, as illustrated by the agreement between the golden circles, calculated with $\lambda=\sqrt{3.}$, $M=20$, and the indicated $\zeta$ and the Keldysh curve.

\section{Conclusions}
\label{sec:conclusions}
We have examined the resonant-level model with a view to addressing a simple question: is the coupling to the conduction band in this uncorrelated Hamiltonian sufficient to destroy the coherence of the transitions from the impurity to the metal? In the continuum ($L\to\infty$) limit, the answer to this question is affirmative. Under a switch-on, the sequence that exhibits the richest physics, the impurity-metal transitions, which drive the system towards equilibrium, constitute the dominant source of decoherence. The Anderson catastrophe is a secondary source that only becomes significant when the coupling is very small. Our computations ratify these notions and give quantitative substance to them.

The time dependence of the fidelity in a finite system, with large lattice size $L$, is virtually identical to the continuum limit for a period that is short in comparison with the transit time, that is, the time required by the perturbation created by the activation of the coupling to reach the opposite end of the lattice. In our units, that time is equal to the lattice size $L$.  At the end of that period, well before $t=\tfrac{L}2$ significant deviations become evident in the time dependence of the fidelity. Starting at $t=\frac{L}2$, as expected, the time evolution is reversed, evidencing the revival of coherence.

The real-space numerical renormalization-group (eNRG) construction described in Sec.~\ref{sec:enrg-method} is immune to this undesirable finite-size effect, which limits the accuracy of numerical computations. As illustrated by the comparison with the benchmark in Fig.~\ref{fig:5}, the eNRG results for $\ff(t)$ are essentially exact over the entire interval $0<t<\zeta$, hence offering a reliable description of the continuum limit.

Another limitation of the straightfoward diagonalization of the Hamilstonian~\eqref{eq:1} becomes apparent as $\Gamma$ is reduced, so that the impurity width becomes comparable to the minimum interval $\Delta\epsilon$ in the spectrum of the tight-binding Hamiltonian. As the plots in Figs.~\ref{fig:7} show, the calculated properties then display deviations that depend on the ratio $\Delta\epsilon\Gamma$ and are, therefore, $L$-dependent. The example in the same figure shows that the eNRG data are much less sensitive to this size-effect.

Other questions remain unanswered: how would correlation affect our conclusions? Given that our study has been restricted to small couplings, would our findings be the same if $\Gamma$ were larger than $\tau$? Under very strong coupling, the impurity couples with a localized orbital in the metal to form a bonding-antibonding pair that is approximately decoupled from the metal. This special limit deserves special attention and will be the subject of future studies. 

The results in this paper highlight the convenience of the eNRG procedure to calculate time-dependent properties. Our runs explored the flexibility provided by the parameter $\zeta$ to examine the different aspects of the study and to extract physical insight from results that covered relatively long time intervals. As Sec.~\ref{sec:coher-tight-bind} demonstrated, the eNRG procedure is immune to the commensurability problem. It yields accurate results even in time ranges when this limitation makes it unreliable to compute physical properties by directly diagonalizing a tight-binding Hamiltonian.

But now it is time to leave these technical issues aside and thank Amir for his gigantic contribution to our community. As a teacher, a
source of inspiration, a driver of enthusiasm, and a great friend, he is unrivaled. One of us has
known him for more than four decades and cannot count the blessings that have come from his
personality, directly or through his remarkable scientific genealogy. Keep going strong, Professor Caldeira, 
and keep us happy!

\section*{ Acknowledgments}
We acknowledge the support of the INCT project Advanced Quantum
Materials, involving the Brazilian agencies CNPq (grant 408766/2024-7), FAPESP (2025/27091-3) and
CAPES. We also acknowledge the support of FAPESP (grant 2022/05198-2), CNPq (311689/2023-0), and
the Coordenação de Aperfeiçoamento de Pessoal de N ível Superior – Brasil (CAPES) – Finance Code
001. The work of FDP was supported by PhD and internship fellowships from FAPESP ( grants 2022/09312-4
and 2024/05637-1).

\bibliography{2024boxReference,bibliography}
\end{document}